\documentclass{aa}
\usepackage{epsfig}
\usepackage{graphics}
\usepackage{longtable}
\begin{document}

\title{A study of the \ion{Mg}{ii} 2796.34 \AA\ emission line in
late--type normal, and RS CVn stars}

\author{Daniela Cardini\inst{1}
\and Angelo Cassatella\inst{1}\inst{,2}
\and Massimo Badiali\inst{1}
\and Aldo Altamore\inst{2}
\and Maria--Jos\'e Fern\'andez--Figueroa\inst{3}
}

\offprints{Angelo Cassatella,
\email{cassatella@fis.uniroma3.it}}

\institute {Istituto di Astrofisica Spaziale e Fisica Cosmica, CNR,
Via del Fosso del Cavaliere 100, 00133 Roma, Italy \and Dipartimento
di Fisica E. Amaldi, Universit\`a degli Studi Roma Tre, Via della
Vasca Navale 84, 00146 Roma, Italy \and Departamento de
Astrof\'{\i}sica, Facultad de F\'{\i}sica, Universidad Complutense,
28040 Madrid, Spain }

\date{Received / Accepted}

\authorrunning{D. Cardini et al.}

\titlerunning{Flux--width relationship in the \ion{Mg}{ii} 2796.34 \AA\
emission}

\abstract{We carry out an analysis of the \ion{Mg}{ii} 2796.34 \AA\
emission line in RS CVn stars and make a comparison with the normal
stars studied in a previous paper (Paper I).  The sample of RS CVn
stars consists of 34 objects with known {\it HIPPARCOS} parallaxes and
observed at high resolution with {\it IUE}.  We confirm that RS CVn
stars tend to possess wider \ion{Mg}{ii} lines than normal stars
having the same absolute visual magnitude.  However, we could not find
any correlation between the logarithmic line width $log~W_{\circ}$ and the
absolute visual magnitude $M_V$ (the Wilson--Bappu relationship) for
these active stars, contrary to the case of normal stars addressed in
Paper I. On the contrary, we find that a strong correlation exists in
the ($M_V$, $log~L_{\ion{Mg}{ii}}$) plane ($L_{\ion{Mg}{ii}}$ is the
absolute flux in the line).  In this plane, normal and RS CVn stars
are distributed along two nearly parallel straight lines with RS CVn
stars being systematically brighter by ${\approx}$ 1 dex.  Such
a diagram provides an interesting tool to discriminate active from
normal stars. We finally analyse the distribution of RS CVn and of
normal stars in the ($log~L_{\ion{Mg}{ii}}$, $log~W_{\circ}$) plane, and
find a strong linear correlation for normal stars, which can be used
for distance determinations.

 }

\maketitle

\begin{table*}
\caption{Basic parameters of the RS CVn stars in the sample}
\begin{flushleft}
\begin{tabular}{@{} r l l r r r r  l r}
\hline \hline \noalign{\smallskip}

HIP&Name&$Sp. Type$ & $P$(days) &$B-V$&$\pi$(mas) & $ M_{\it V}$& $\Delta M_{\it V}$ &  $Active~Star$ \\

\hline \noalign{\smallskip}
  4157 & CF Tuc       & G0 V + K4 IV      &   2.8 $^2$&   0.68 & 11.60$\pm$ 0.65 &   2.93 &   0.36  & Cool$^a$        \\
  9630 & XX Tri       & K0 III            &  24.3 $^2$&   1.11 &  5.08$\pm$ 1.10 &   1.92 &   0.36  &                \\
 10280 & TZ Tri       & F5 + K0 III       &  14.7 $^2$&   0.77 & 10.68$\pm$ 0.92 &   0.08 &   0.06  & Cool$^e$       \\
 13118 & VY Ari       & K3-4 V-IV         &  16.6 $^1$&   0.96 & 22.73$\pm$ 0.89 &   3.72 &   0.16  &                \\
 16042 & UX Ari       & G5 V + K0 IV      &   6.4 $^1$&   0.88 & 19.91$\pm$ 1.25 &   2.96 &   0.04  & Cool$^f$       \\
 16846 & V711 Tau     & G5 V-IV + K1 IV   &   2.8 $^2$&   0.88 & 34.52$\pm$ 0.86 &   3.51 &   0.09  & Both$^d$        \\
 19431 & EI Eri       & G5 IV             &   1.9 $^2$&   0.71 & 17.80$\pm$ 0.96 &   3.28 &   0.10  &              \\
 23743 & BM Cam       & K0 III            &  80.9 $^1$&   1.11 &  5.22$\pm$ 0.92 &  -0.33 &   0.08  &              \\
 24608 & $\alpha$ Aur & G1 III + K0 III   &  8/80 $^2$&   0.80 & 77.29$\pm$ 0.93 &  -0.48 &   0.04  & Hot$^a$        \\
 35600 & AR Mon       & G8 III + K2-3 III &  21.2 $^3$&   1.06 &  3.62$\pm$ 1.22 &   1.52 &   0.91  & Both$^d$        \\
 37629 & $\sigma$ Gem & K1 III            &  19.4 $^2$&   1.12 & 26.68$\pm$ 0.80 &   1.36 &   0.07  &              \\
 46159 & IL Hya       & G8 V + K0 III-IV  &  12.8 $^1$&   1.01 &  8.36$\pm$ 0.86 &   1.96 &   0.19  & Cool$^i$        \\
 56851 & V829 Cen     & G5 V + K1 IV      &  11.7 $^2$&   0.95 &  8.22$\pm$ 0.83 &   2.44 &   0.20  &              \\
 59600 & HU Vir       & K0 III-IV         &  10.3 $^2$&   0.97 &  8.00$\pm$ 1.25 &   3.21 &   0.18  &              \\
 59796 & DK Dra       & K1 III + K1 III   &  63.0 $^1$&   1.15 &  7.24$\pm$ 0.55 &   0.59 &   0.14  & Both$^a$        \\
 64293 & RS CVn       & F4 IV + G9 IV     &   4.8 $^2$&   0.61 &  9.25$\pm$ 1.06 &   2.90 &   1.27  & Cool$^b$       \\
 65187 & BM CVn       & K1 III            &  20.6 $^2$&   1.16 &  9.00$\pm$ 0.76 &   2.08 &   0.13  &              \\
 81519 & WW Dra       & G2 IV + K0 IV     &   4.6 $^2$&   0.71 &  8.67$\pm$ 1.24 &   2.92 &   0.65  & Both$^c$       \\
 82080 & $\varepsilon$ UMi & A8-F0 V +G5 III& 39.5 $^3$&   0.90 &  9.41$\pm$ 0.67 &  -0.92&   0.06  & Cool$^c$         \\
 84586 & V824 Ara     & G5 IV + K0 V-IV   &   1.7 $^2$&   0.80 & 31.83$\pm$ 0.73 &   4.38 &   0.12  & Both$^a$       \\
 85852 & DR Dra       & WD + K0-2 III     &  31.5 $^2$&   1.04 &  9.68$\pm$ 0.80 &   1.54 &   0.06  &             \\
 87965 & Z Her        & F4 V-IV + K0 IV   &   4.0 $^1$&   0.60 & 10.17$\pm$ 0.84 &   2.28 &   0.82  & Cool$^b$       \\
 88848 & V815 Her     & G5 V              &   1.8 $^2$&   0.73 & 30.69$\pm$ 2.09 &   5.13 &   0.12  &              \\
 94013 & V1762 Cyg    & K1 IV-III         &  28.6 $^2$&   1.09 & 14.24$\pm$ 0.48 &   1.65 &   0.09  &              \\
 95244 & V4138 Sgr    & K1 III            &  60.2 $^2$&   1.03 & 11.40$\pm$ 0.86 &   2.00 &   0.17  &             \\
 96003 & V1817 Cyg    & A2 V + K2 III-IV  & 108.8 $^2$&   1.12 &  3.10$\pm$ 0.50 &  -1.17 &   0.06  &                \\
 96467 & V1764 Cyg    & F-K1 III:         &  40.2 $^1$&   1.22 &  3.44$\pm$ 0.90 &   0.45 &   0.16  &                \\
109002 & HK Lac       & F1 IV + K0 III    &  24.4 $^1$&   1.05 &  6.62$\pm$ 0.61 &   1.02 &   0.22  & Cool$^g$       \\
109303 & AR Lac       & G2 IV + K0 IV     &   2.0 $^1$&   0.76 & 23.79$\pm$ 0.59 &   2.99 &   0.74  & Both$^h$      \\
111072 & V350 Lac     & K2 III            &  17.7 $^1$&   1.17 &  8.18$\pm$ 0.56 &   0.97 &   0.10  &                \\
112997 & IM Peg       & K2 III-II         &  24.4 $^2$&   1.13 & 10.33$\pm$ 0.76 &   0.93 &   0.12  &                \\
114639 & SZ Psc       & F8 V + K1 IV      &   4.0 $^2$&   0.79 & 11.34$\pm$ 0.92 &   2.67 &   0.49  & Cool$^c$       \\
116584 & $\lambda$ And & G8 III-IV        &  54.3 $^1$&   0.98 & 38.74$\pm$ 0.70 &   1.75 &   0.19  &                \\
117915 & II Peg       & K2-3 V-IV         &   6.7 $^1$&   1.01 & 23.62$\pm$ 0.90 &   4.38 &   0.23  &                \\
\noalign{\smallskip} \hline \noalign{\smallskip}
\end{tabular}
\par
\vspace{0.3cm} \noindent {\it Notes}: { Spectral types are from the
Catalog of Active Binary Stars (CABS) by Strassmeier et al. (1993)
except for HU Vir (see Fekel et al. 1999; Montes et al. 2000).  The
$B-V$ colour the parallax $\pi$, the absolute visual magnitude $M_{\it
V}$, and the amplitude of its variations $\Delta M_{\it V}$ are from
the {\it HIPPARCOS} Catalogue and its Variability Annex.\\ The
photometric periods $P$ are indicated in Col. 4 together with the
corresponding reference: 1) Variability Annex to the Hipparcos
Catalogue; 2) CABS; 3) Hipparcos Input Catalogue.\\ For objects with
double--lined spectra, the last column indicates which one of the
stellar components is the active star, according to the following
sources: a) Gunn et al. (1998); b) Fern\'andez--Figueroa et al. 1986;
c) Fern\'andez--Figueroa et al. 1994; d) Montes et al. (1995a); e)
Montes et al. (1995b); f) Montes et al. (1996); g) Montes et
al. (1997); h) Pagano et al. (2001); i) Fekel et al. (1999) }
\label{tab:data1}
\end{flushleft}
\end{table*}

\section{Introduction}
\label{sec:intro}

The relationship between the absolute visual magnitude $M_V$ and the
width of the optical \ion{Ca}{ii} K emission line was discovered by
Wilson \& Bappu (1957) for stars with spectral type later than G0, and
is commonly designed as the Wilson--Bappu effect (WB).  This
relationship has been widely studied in the optical (see
\cite{wallerstein}, and references therein).  In more recent years it
has been shown that a similar relationship applies also to the
\ion{Mg}{ii} k emission line at 2796.34 \AA, which has the same
chromospheric origin as the \ion{Ca}{ii} K line (see \cite{mclintock},
\cite{garcia}, \cite{vladilo}, \cite{elgaroy3}). As for the
theoretical interpretation of the Wilson--Bappu effect see Gayley
(2002), the comprehensive review by Linsky (1999), and references
therein.  The unprecedented accuracy of the {\it HIPPARCOS} parallax
determinations (ESA 1997) led to a substantial upgrading of the WB
relationship, both in the optical (\cite{wallerstein}) and in the
ultraviolet ranges (\cite{scoville}, \cite{elgaroy3}, and
\cite{cassat} -- hereafter Paper I).

Active stars and, in particular, RS CVn stars were intentionally
excluded in most of the above studies because of their binary nature
and their enhanced chromospheric activity, but were however object of
specific investigations based on the emission doublets
from \ion{Ca}{ii} (Montes et al.  1994) or \ion{Mg}{ii} (Elgar{\o}y et
al. 1997; \"Ozeren et al.  1999). These studies lead to the definition
of a WB relationship for active stars, different from
that of normal stars.

In this paper we bring new observational results concerning the
behaviour of the width and luminosity of the \ion{Mg}{ii} k line
in RS CVn and normal stars.  Our specific purposes are:\\

\noindent a) To verify to which extent the observed \ion{Mg}{ii} k
line width is correlated with the absolute visual magnitude for RS CVn
stars, if proper account is given for their intrinsic variability.


\noindent b) To quantify the enhancement of the chromospheric
activity in RS CVn stars with respect to normal stars on the basis of
the intrinsic \ion{Mg}{ii} k line luminosity. \\

\noindent c) To define  a \ion{Mg}{ii} k luminosity versus line width
relationship for normal stars, which can be used for distance
determinations.

The sample of RS CVn stars, the observations and the data
reduction are presented in Sections \ref{sec:sample} and
\ref{sec:observa}.  In Section \ref{sec:wbrs} we discuss the
feasibility of defining a WB diagram for RS CVn stars. A comparison of
RS CVn stars with normal stars in the \ion{Mg}{ii} luminosity versus
$M_V$ diagram is provided in Sect. \ref{sec:discriminating}. In
Sect. \ref{sec:lummg2} we present our results about the \ion{Mg}{ii} k
luminosity-- width correlation for normal stars and discuss the case
of RS CVn stars. The conclusions are given in
Sect. \ref{sec:conclusions}.

\begin{table*}
\caption{Results of the {\it IUE} measurements for RS CVn stars}
\begin{flushleft}
\begin{tabular}{@{} r l r r r r r r r r r r}
\hline \hline \noalign{\smallskip}
HIP  & Name  & & $M_{\it V}(IUE)$ & $\Delta M_{\it V}(IUE)$ & $log~W_{\circ}$ & $\Delta log~W_{\circ}$ & & $log~L_{\ion{Mg}{ii}}$ & $\Delta log~L_{\ion{Mg}{ii}}$ &
$N$  \\
\hline \noalign{\smallskip}
  4157 & CF Tuc       &  &   2.76 &   0.17 &   2.16 &   0.15 &  &  30.41 &   0.30 &    8 & \\
  9630 & XX Tri       &  &   1.86 &        &   2.07 &        &  &  31.23 &         &    1 & \\
 10280 & TZ Tri       &  &   0.16 &   0.05 &   2.19 &   0.02 &  &  31.21 &   0.08 &    3 & \\
 13118 & VY Ari       &  &   3.59 &   0.17 &   1.91 &   0.01 &  &  30.13 &   0.06 &    5 & \\
 16042 & UX Ari       &  &   2.89 &   0.43 &   2.09 &   0.30 &  &  30.60 &   0.53 &   65 & \\
 16846 & V711 Tau     &  &   3.45 &   0.46 &   2.10 &   0.48 &  &  30.45 &   0.46 &  264 & \\
 19431 & EI Eri       &  &   3.29 &   0.16 &   2.12 &   0.23 &  &  30.23 &   0.18 &   28 & \\
 23743 & BM Cam       &  &  -0.32 &   0.10 &   2.23 &   0.01 &  &  31.77 &   0.08 &    2 & \\
 24608 & $\alpha$ Aur &  &  -0.45 &   0.22 &   2.16 &   0.23 &  &  31.06 &   0.15 &   76 & \\
 35600 & AR Mon       &  &   1.43 &        &   2.41 &        &  &  30.99 &         &    1 & \\
 37629 & $\sigma$ Gem &  &   1.36 &   0.26 &   2.13 &   0.30 &  &  31.09 &   0.33 &   27 & \\
 46159 & IL Hya       &  &   2.15 &        &   2.02 &        &  &  30.89 &         &    1 & \\
 56851 & V829 Cen     &  &        &        &   1.99 &        &  &  30.67 &         &    1 & \\
 59600 & HU Vir       &  &   3.25 &        &   2.04 &        &  &  30.46 &         &    1 & \\
 59796 & DK Dra       &  &   0.50 &   0.14 &   2.12 &   0.13 &  &  31.40 &   0.14 &    4 & \\
 64293 & RS CVn       &  &   2.98 &   0.14 &   2.05 &   0.05 &  &  29.96 &   0.08 &    3 & \\
 65187 & BM CVn       &  &        &        &   2.06 &        &  &  30.89 &         &    1 & \\
 81519 & WW Dra       &  &   3.10 &        &   2.28 &        &  &  29.97 &         &    1 & \\
 82080 & $\varepsilon$ UMi  &  &  -0.94 &   0.00 &   2.15 &   0.10 &  &  31.39 &   0.08 &    2 & \\
 84586 & V824 Ara     &  &   4.30 &   0.07 &   2.13 &   0.05 &  &  30.05 &   0.33 &    3 & \\
 85852 & DR Dra       &  &   1.62 &        &   2.05 &        &  &  30.88 &         &    1 & \\
 87965 & Z Her        &  &   2.29 &        &   1.82 &        &  &  29.92 &         &    1 & \\
 88848 & V815 Her     &  &   5.10 &   0.05 &   1.87 &   0.16 &  &  29.33 &   0.10 &    4 & \\
 94013 & V1762 Cyg    &  &   1.76 &   0.19 &   2.09 &   0.04 &  &  30.98 &   0.06 &    4 & \\
 95244 & V4138 Sgr    &  &   2.17 &        &   1.92 &        &  &  30.76 &         &    1 & \\
 96003 & V1817 Cyg    &  &  -1.23 &        &   2.22 &        &  &  32.11 &         &    1 & \\
 96467 & V1764 Cyg    &  &   0.35 &        &   2.34 &        &  &  31.53 &         &    1 & \\
109002 & HK Lac       &  &   1.03 &   0.13 &   2.14 &   0.07 &  &  31.36 &   0.20 &    4 & \\
109303 & AR Lac       &  &   3.08 &   0.90 &   2.28 &   0.33 &  &  30.19 &   0.47 &  198 & \\
111072 & V350 Lac     &  &   1.03 &   0.05 &   2.16 &   0.09 &  &  30.97 &   0.11 &    2 & \\
112997 & IM Peg       &  &   0.94 &   0.22 &   2.15 &   0.17 &  &  31.34 &   0.20 &   12 & \\
114639 & SZ Psc       &  &   2.64 &   0.33 &   2.22 &   0.17 &  &  30.54 &   0.23 &   25 & \\
116584 & $\lambda$ And      &  &   1.75 &   0.35 &   2.01 &   0.22 &  &  30.90 &   0.33 &   79 & \\
117915 & II Peg       &  &   4.33 &   0.44 &   1.94 &   0.33 &  &  30.07 &   0.45 &   59 & \\

\noalign{\smallskip} \hline \noalign{\smallskip}
\end{tabular}
\par
\vspace{0.3cm}

\noindent {\it Notes}: 
$M_{\it V}(IUE)$ and $\Delta M_{\it V}(IUE)$
are the mean visual absolute magnitude and its maximum amplitude. $log
W_{\circ}$ and $\Delta log W_{\circ}$ are, respectively, the logarithm of the mean
full-width at half-maximum for the $\ion{Mg}{ii}$ line and the
corresponding spread.  $log~L_{\ion{Mg}{ii}}$ and $\Delta
log~L_{\ion{Mg}{ii}}$ are, respectively, the logarithm of the mean
luminosity of the \ion{Mg}{ii} line and its spread. $N$ is the number
of the spectra used for each star. Units for the line widths $W_{\circ}$ and
\ion{Mg}{ii} luminosities are in km~s$^{-1}$ and erg~s$^{-1}$,
respectively
\label{tab:data2}
\end{flushleft}
\end{table*}

\section{The present sample of stars}
\label{sec:sample}

We have searched in the Strasbourg Data Center (SIMBAD data base) for
chromospherically active binary stars of the RS CVn type for which
{\it HIPPARCOS} parallax determinations and {\it IUE} long--wavelength
high resolution spectra were available. The search lead to a total of
55 stars. Out of these, 12 were rejected because the {\it IUE} data
were underexposed or saturated and additional six for not fulfilling
the criteria which define the RS CVn class of stars (Fekel et
al. 1986, Montesinos et al. 1994): one for being a pulsating star
(HIP77512, see Fernie 1999), and five for not having been detected as
variable stars by {\it HIPPARCOS}. Out of these latter, HIP39348 and
HIP107095 only show micro variability, while HIP57565, HIP66257 and
HIP60582 are not variable according to the Variability Annex to the
Catalogue. HIP44164 (TY Pyx) was excluded in spite of being an RS CVn
star because the \ion{Mg}{ii} k emission would need a dedicated
analysis due to the complexity of its profile, in which the individual
contribution from the components of the binary system is in general
very evident. Finally, we have rejected V368 Cep and EP Eri because
they are not RS CVn stars: V368 Cep has been identified as a member of
the Local Association and is classified as a post T Tauri star, and EP
Eri, is also a nearby young star (see Montes et al. 2001, and
references therein). The final sample of RS CVn stars consists then
of 34 stars.

The relevant information concerning the selected objects is given
in Table \ref{tab:data1}, which provides in Columns 1 to 4: the {\it
HIPPARCOS} number, the star name, the spectral type and the
photometric period. Columns 5 to 8 provide the $B-V$ colour index, the
parallax with its error, the absolute magnitude $M_{V}$, and the
maximum amplitude of the variation in the visual band ${\Delta}
M_{V}$. Finally, the last Column specifies which  component of
the binary system is the active star. The relevant references to
the above data are specified as footnotes to the Table.

\section{Observations and data reduction}
\label{sec:observa}
The {\it IUE} high resolution long wavelength (LWP and LWR) spectra
were obtained from the {\it INES} (IUE Newly Extracted Spectra)
retrieval system through its Principal Centre at {\tt http:
//ines.vilspa.esa.es}.  A full description of the {\it INES} high
resolution spectra is given in Cassatella et al. (2000),
Gonz\'alez-Riestra et al. (2000) and Gonz\'alez-Riestra et al. (2001).
The spectra were inspected individually in the \ion{Mg}{ii} 2800 \AA\
wavelength region in order to reject noisy, overexposed or
underexposed data, as in Paper I.  The total number of spectra used
for this investigation is 889.  The \ion{Mg}{ii} k line width and flux
in individual spectra were measured using the procedures described
below. The results of the measurements are given in Table
\ref{tab:data2} and discussed later in detail.

Note that a considerable fraction of the observations is concentrated
on a few objects and especially  on V711 Tau (264 spectra), AR Lac (198
spectra), ${\lambda}$ And (79 spectra), and ${\alpha}$ Aur (76 spectra).
The number of spectra for each star is specified in the last column of
Table \ref{tab:data2}.

\subsection{The \ion{Mg}{ii} k line width}

The \ion{Mg}{ii} k line widths were determined through the procedure
described  in Paper I, which consists, as a first step, in
defining the start and end wavelengths of the violet and  red
wings of the \ion{Mg}{ii} k profile and, as a second step, in fitting
together the two selected regions with an unique Gaussian profile. In
this way a portion of the profile can be excluded, whenever needed.

The line widths were defined as the full width of the {\it fitted}
profile measured at half maximum of the {\it observed} profile.

The reason for adopting this procedure instead of measuring the line
width directly on the observed line profiles, is that it allows accurate
measurements to be made also in presence of noticeably asymmetric
profiles, or profiles having a central depression such that the peak
intensity of one of the two wings is less than 50\% of the other.  It
is clear that, in normal stars, the central depression is essentially
due to self--reversal or interstellar/circumstellar absorption.

In our specific case, however, an ``absorption'' dip in the profile or
the presence of profile asymmetries may represent a signature of the
binary nature of the object under study. A clear example of this case
is the RS CVn eclipsing binary star AR Lac: the detailed study of
Pagano et al. (2001) shows that the profile changes of the
\ion{Mg}{ii} doublet are correlated with the orbital period so that
the wavelength separation of the \ion{Mg}{ii} k emission from the K0
IV and the G2 IV components is largest at orbital quadrature,
i.e. when the Doppler shift is maximum (see also
Sect. \ref{sec:wbrs}).  In order to preserve the statistical character
of this investigation, no attempt will be made to fit the \ion{Mg}{ii}
k line with two emission components as done for AR Lac by Pagano et
al. (2001). In any case, AR Lac is, together with TY Pyx (not treated
here; see Sect. \ref{sec:sample}), the only target in our sample in
which the contribution from the two stars can be separated as a
consequence of their similar \ion{Mg}{ii} luminosity and of the
favorable inclination angle of the system.

Before performing the Gaussian fit, all spectra were re--sampled to
increase the number of data points by a factor 5, and smoothed.

The errors on measured line widths were estimated to be equal to half
of the sampling interval of the spectra, corresponding to 3.64
km~s$^{-1}$.  The measured line widths were corrected for instrumental
broadening according the following relation:

\begin{equation}
\rm{W_{\circ}^2 = {W}^2 -b^2}
\label{eq:corr}
\end{equation}

\noindent where $W$ is the measured width, $W_{\circ}$ the corrected value
and $ b=18~ km~s^{-1}$ is the FWHM of the {\it IUE} Point Spread
Function for high resolution spectra, assumed to be Gaussian
(\cite{evans}).  The logarithm of the mean line width for each star,
$log~W_{\circ}$, is reported in Column 5 of Table \ref{tab:data2}.  Whenever
more than one spectrum was available we have reported in Column 6 the
maximum amplitude of the variations, ${\Delta log~W_{\circ}}$.

\subsection{The \ion{Mg}{ii} k line fluxes}
\label{sec:mg2fluxes}

The \ion{Mg}{ii} k line fluxes for the RS CVn stars were obtained by
direct integration of the observed profiles, as measured above the
underlying local continuum.  No attempt has been made to correct the
\ion{Mg}{ii} k flux for interstellar absorption since we are dealing
with nearby stars and these corrections are small.  Fluxes at the
Earth were converted into \ion{Mg}{ii} k absolute luminosity using the
parallaxes in Table \ref{tab:data1}; the logarithm of their mean
values, $log~ L_{\ion{Mg}{ii}}$, are reported in Table
\ref{tab:data2}. Whenever more observations were available, the
maximum amplitude of luminosity variations $\Delta
log~L_{\ion{Mg}{ii}}$ is also given.


As for the errors on the absolute line luminosity, these were derived
from the errors on the parallaxes and assuming an estimated constant
error of 15 \% on the measured fluxes.

For comparison purposes, the \ion{Mg}{ii} k flux measurements were
also extended to the normal stars studied in Paper I. Their values
will be used later (see Fig. \ref{fig:magflunor}a).

\subsection{Absolute visual magnitudes }
\label{sec:mag}

We have evaluated the visual magnitudes $V_{IUE}$ from the counts
of the Fine Error Sensor (FES) on board {\it IUE}, taken in
correspondence to the spectroscopic observations.

The procedure used to obtain the FES visual magnitudes consists in
correcting the FES counts for the dead time and for the
time--dependent sensitivity degradation, in converting the corrected
counts into FES magnitudes, and in reporting them into the Johnson
photometric system through the application of a colour correction
which depends on the star's colour index B-V.

Several FES calibrations have been published during the {\it IUE}
lifetime. They mainly differ for the way the FES time--dependent
degradation is corrected for (see Barylak et al. 1985, Imhoff \&
Wasatonic 1986, Fireman \& Imhoff 1989, P\'erez \& Loomis 1991, and
P\'erez 1992).  For the observations obtained before 1990, we adopted
the correction algorithm suggested by P\'erez \& Loomis (1991), which
is based on an accurate analysis of a very large number of
observations. For dates after 1990 (date of the change of the FES
reference point), we used the algorithm by \cite{perez}.

The absolute magnitudes were computed from the {\it HIPPARCOS}
trigonometric parallaxes without any allowance for interstellar
reddening.  The corresponding errors were derived from the errors on
the trigonometric parallaxes and assuming a typical error of $\pm$
0.01 mag on the visual magnitudes from the {\it HIPPARCOS} Catalogue,
and ${\pm}$ 0.05 mag for $V_{\it IUE}$ as suggested by P\'erez \&
Loomis (1991).

The mean values of the absolute magnitudes ${M_{V}(IUE)}$ for the RS
CVn stars in our sample and the maximum amplitude of the variations
${\Delta M_{V}(IUE)}$ are reported in Table \ref{tab:data2}, except
for HIP56851 and HIP65187 (no FES observations were available).  

We note that the amplitudes of the photometric variations ${\Delta
M_{V}(IUE)}$ do not coincide exactly with the corresponding values
from the {\it HIPPARCOS}~ Catalogue (cf. Col. 8 and 4 of Tables 1 and
2, respectively). This is due to the considerably longer time span
covered by the {\it HIPPARCOS} observations and to the different kind
of monitoring.

\begin{figure}
\begin{center}
\includegraphics[width=8.8cm]{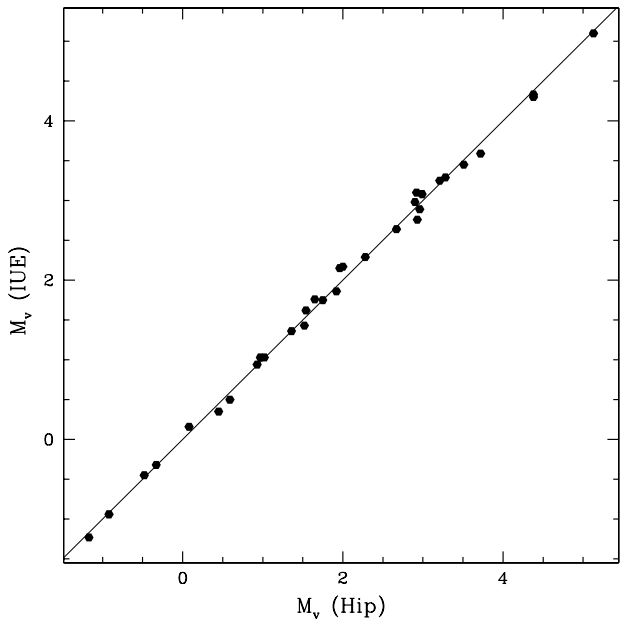}
\caption{The mean absolute visual magnitude derived from the {\it IUE}
FES counts, $M_V(IUE)$, is plotted against the {\it HIPPARCOS} absolute
visual magnitude $M_V$ for the stars in Tables \ref{tab:data1} and
\ref{tab:data2}. The figure shows that there are no systematic errors
when using {\it IUE} magnitudes over the wide range of values covered}
\label{fig:magmag}
\end{center}
\end{figure}

In Fig. \ref{fig:magmag} we show a comparison of the absolute visual
magnitudes obtained from the FES counts and the corresponding {\it
HIPPARCOS} values.  The two sets of data agree to within 0.08 mag
(rms). Since this value is only slightly larger than the expected error
on $V_{\it IUE}$, and given the statistical nature of this paper, we
will  use as a default the {\it HIPPARCOS} values, without any
significant loss of accuracy. The ${M_{V}(IUE)}$ values will be used
only in the specific case of AR Lac, discussed in the next Section.

\section{The WB diagram for RS CVn stars}
\label{sec:wbrs}

\begin{figure}
\includegraphics[width=8.8cm]{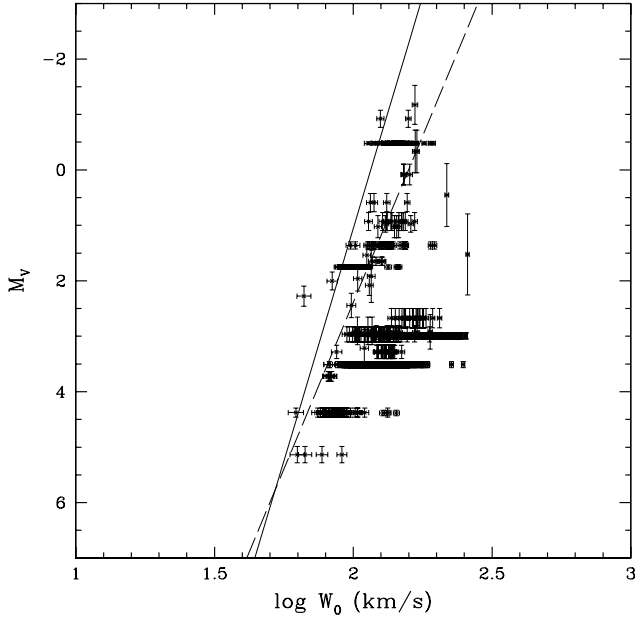}
\caption{The Wilson--Bappu effect in the \ion{Mg}{ii} k line for RS
CVn stars. Overlayed to the data points we show the WB relationship
for normal stars form Paper I (full line), and for RS CVn stars
(dashed line)  from \"Ozeren et al. (1999) }
\label{fig:cercastella}
\end{figure}
 
\begin{figure}

\includegraphics[width=8.8cm]{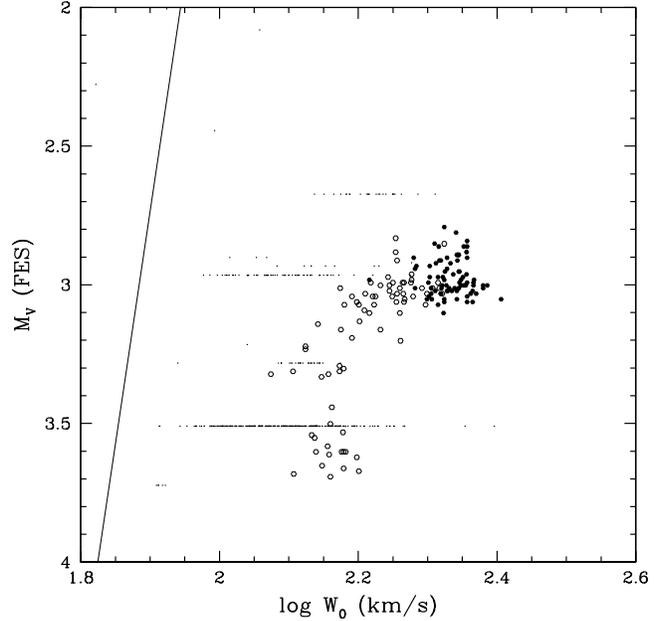}
\caption{The Wilson--Bappu effect in a restricted range of
coordinate's values to show the large variations of AR Lac
(circles). Filled circles correspond to data close to quadrature (see
text).  The straight line represents the relationship for normal stars
described in Paper I (see Eq. \ref{eq:bappueq}). Dots represent the
other RS CVn stars}
\label{fig:arlac}
\end{figure}

{As mentioned in the Introduction, the Wilson-Bappu effect in
active stars has been studied by Montes et al. (1994), Elgar{\o}y et
al. (1997), and \"Ozeren et al. (1999).  These authors found that
active stars, and RS CVn stars in particular, have substantially
broader lines than normal stars and that, in spite of the large
spread, the line widths are well correlated with the visual absolute
magnitudes (r=0.80 to 0.93).  We wish to stress that in all the above
analyses, the mean values for the line widths were used instead of the
individual determinations.

On the other hand, we believe that the variety of orbital and
structural parameters in these binary systems should have important
effects on the observed line widths. For this reason we have decided
to used {\it all} good quality {\it IUE} spectra available for each of
the stars in our sample.  

Our results concerning the Wilson--Bappu effect in RS CVn stars are
shown in Fig. \ref{fig:cercastella}, where the absolute magnitude
$M_V$ is plotted as a function of $log~W_{\circ}$.  The figure reports the
individual measurements together with the corresponding error bars and
indicates, for reference, the regression line describing
the Wilson--Bappu effect for normal stars from Paper I (full line)
 
\begin{equation}
\it {M_{V} = (34.56 {\pm} 0.29)- (16.75 {\pm} 0.14)~ log W_{\circ}}
\label{eq:bappueq}
\end{equation}

\noindent where $W_{\circ}$ is in km~s$^{-1}$. 

In Fig. \ref{fig:cercastella} we also indicate, as a dashed line, the
magnitude--width linear relationship for RS CVn stars by \"Ozeren et
al. (1999).  We cannot make a direct comparison with
other studies because they are based on a different line
(\ion{Ca}{ii}; see Montes et al. 1994), or make use of a different
definition of the \ion{Mg}{ii} line width (Elgar{\o}y et al. 1997).

Fig. \ref{fig:cercastella} shows that RS CVn stars have in general
broader lines than normal stars, as also found by the previously
quoted authors.  It also shows, however, that the magnitude--width
relationship by \"Ozeren et al. (1999) does not provide a good
representation for the stars in our sample.  This is not surprising,
since $M_V$ and $log~W_{\circ}$ in our data set are clearly uncorrelated:
the correlation coefficient for these individual determinations is
indeed as low as r=0.17.  A major cause for the above discrepancy and
for the lack of correlation is most likely the intrinsic variability
and the variety of orbital and structural parameters in these binary
systems.

To underline the influence of orbital motion we consider the
case of AR Lac, which is the second best monitored RS CVn star, with a
total of 198 {\it IUE} observations along 15 years.  AR Lac is a G2 IV
+ K0 IV eclipsing binary system, whose components are spin--orbit
coupled with a 1.98 day period (see Pagano et al. 2001 and references
therein). Given the quite large amplitude of variability for this
object (both in $M_V$ and in $log~W_{\circ}$, see Table \ref{tab:data2}) we
have in this special case computed the absolute magnitudes from the
FES counts obtained at the moment of the spectroscopic observations
and from the {\it HIPPARCOS} parallax (see Sec. \ref{sec:mag}).  A
magnification of the WB diagram, centered around AR Lac is shown in
Fig. \ref{fig:arlac}, without the error bars. This star, indicated
with filled and open circles, describes half a loop in the diagram,
with a considerable excursion in both axes. To understand this
behaviour we have indicated with filled circles the data obtained at
phases ${\phi}$ near quadrature (0.12$\leq$${\phi}$$\leq$0.37 and
0.62$\leq$${\phi}$$\leq$0.88). It is immediately clear that these
phases are characterized by the broadest \ion{Mg}{ii} line profiles
and the highest luminosity levels, which is just what is expected
considering the eclipsing binary nature of AR Lac and the similar V
magnitudes and spectral types of the components of the system.

AR Lac is a limiting case. More in general, our data indicate
that whenever a star has been observed frequently, the fluctuations in
line width ${\Delta log~W_{\circ}}$ may be quite large (10\% to 20\%, see
Table \ref{tab:data2}).

In conclusion, it is not possible with the present data to define a WB
relationship for RS CVn stars because of the impossibility to
disentangle the effects of stellar activity from orbital effects.

\section{The  \ion{Mg}{ii} luminosity vs. $M_V$ relationship}
\label{sec:discriminating}

\begin{figure}
\includegraphics[width=8.8cm]{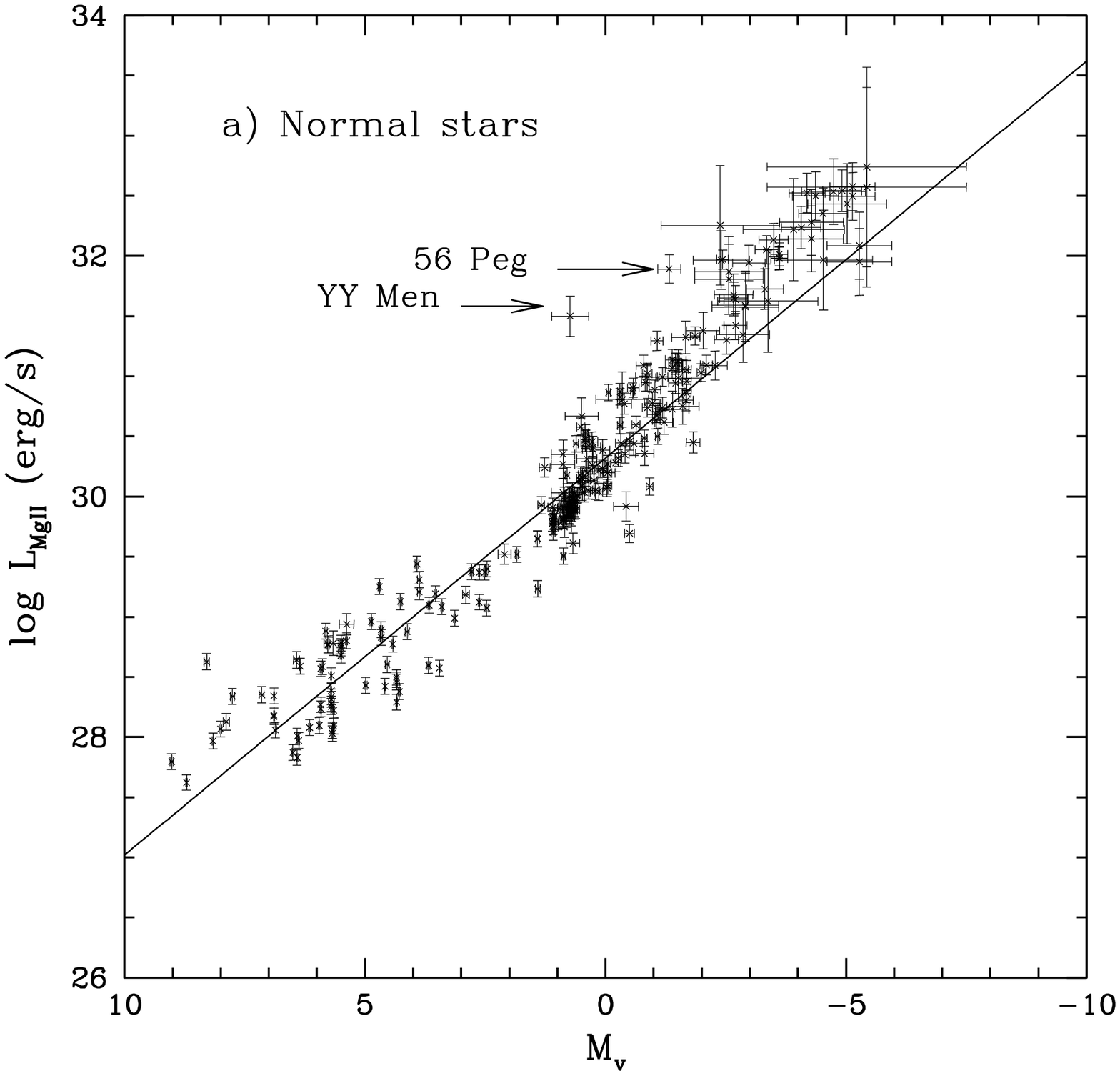}
\includegraphics[width=8.8cm]{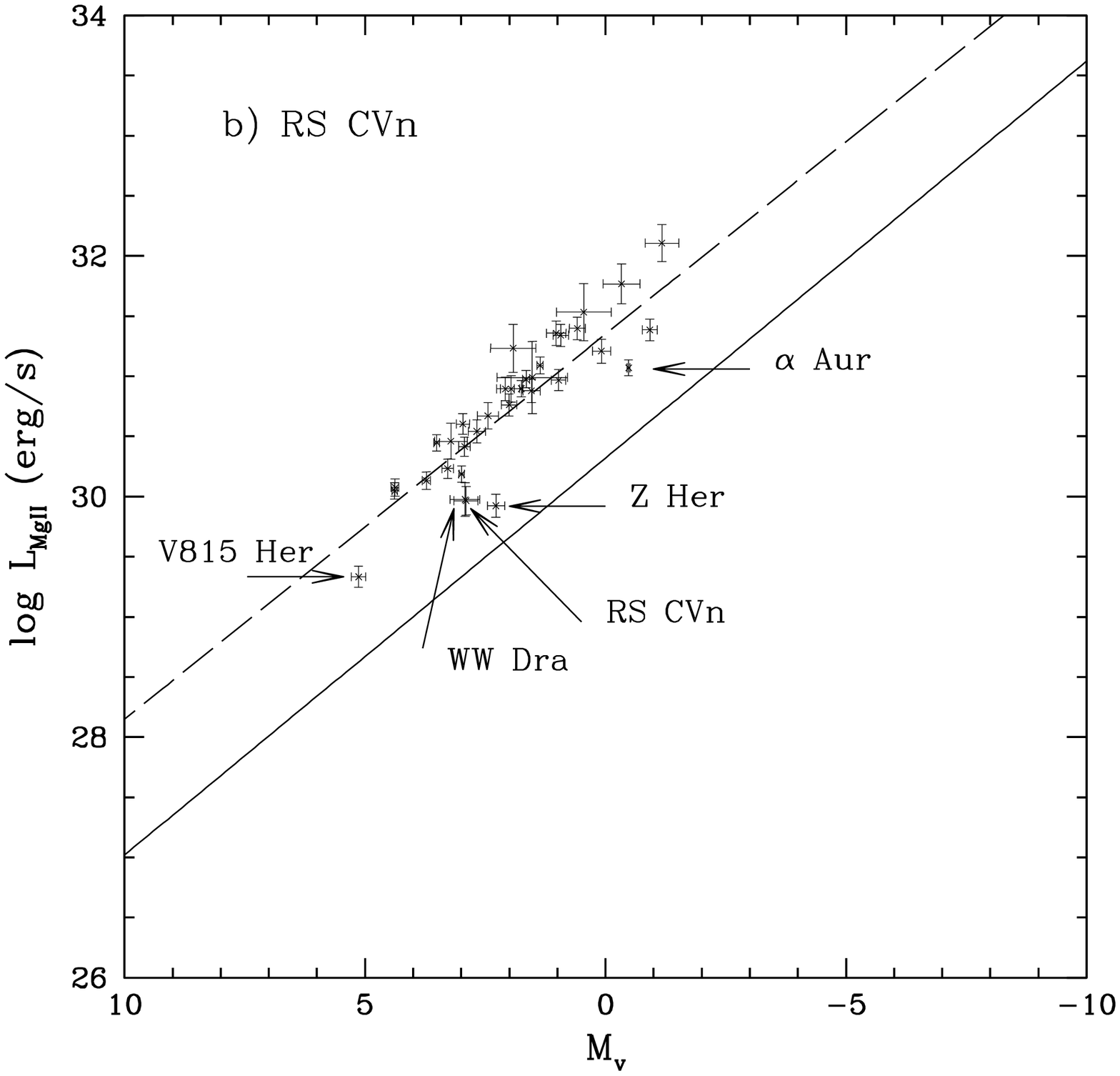}

\caption{ The logarithmic absolute luminosity in the \ion{Mg}{ii} k
line is plotted as a function of the absolute visual magnitude.
\noindent (a) Normal stars. The individual measurements with the
corresponding error bars are indicated. The straight line is the linear fit
to the data (Eq. \ref{eq:lum-mv-normal}).
\noindent (b) RS CVn stars.  The mean values and the corresponding
error bars are reported.  We indicate also the linear fits to the RS
CVn data (dashed line; see Eq. \ref{eq:lum-mv-rs}) and   to the normal
stars  data (full line)}
\label{fig:magflunor}
\end{figure}

As shown by several authors (see e.g. Basri 1987, Cerruti--Sola et
al. 1992, Elgaroy et al. 1997), the \ion{Mg}{ii} k flux is a strong
indicator of stellar activity.  In this Section we consider the
\ion{Mg}{ii} k absolute line luminosity $L_{\ion{Mg}{ii}}$ as an
indicator of the  enhanced activity level of RS CVn stars 
compared with normal stars (see Paper I).
  
The absolute \ion{Mg}{ii} k luminosities were obtained
from the observed fluxes as described in Sect. \ref{sec:mg2fluxes}.

\subsection{Normal Stars}
\label{sec:normalstars}

In Fig. \ref{fig:magflunor}a we show the $log~L_{ \ion{Mg}{ii} }$ vs.
$M_V$ diagram for normal stars. A linear fit to the data, performed by
taking into account the observational errors on both variables,
provides:

\begin{equation}
\it {log~L_{\ion{Mg}{ii}} = (30.32 {\pm} 0.01)- (0.33 {\pm} 0.01)~ M_V} 
\label{eq:lum-mv-normal}
\end{equation}

\noindent where $L_{ \ion{Mg}{ii} }$ is in erg~$s^{-1}$.
In spite of the  wide range of luminosity covered, the
correlation index is very good (r=0.96).

A similar relationship has been obtained by Weiler \& Oegerle (1979)
from {\it OAO 3} moderate--resolution (0.51 \AA) observations of 73
late--type stars.  The difference in the coefficients might be
ascribed to the different resolution and accuracy level of the two
space experiments or, more likely, to the mixture of quiet and active
stars in their sample.

It is worth noticing that two stars, 56 Peg and YY Men,
indicated with arrows in Fig. \ref{fig:magflunor}a, which appear
significantly brighter in the \ion{Mg}{ii} line than predicted on the
basis of the above relationship, are peculiar stars that were
erroneously included in our sample of Paper I. In particular, 56 Peg
is a G8 Ib barium star, reported by Cornide et al. (1992) as the most
active of the sample of 10 barium stars they studied in the
\ion{Ca}{ii} line.  YY Men is a K1 IIIp star belonging to the FK Com
class, as defined by Bopp and Stencel (1981). As evolved
late--type stars, the enhanced chromospheric activity level denoted
by the strength of the \ion{Mg}{ii} k line is likely linked to the
fact that both stars are fast rotators (the rotational period of YY
Men is 9.5476 days according to Cutispoto et al. 1992).

\subsection{RS CVn Stars}
In Fig. \ref{fig:magflunor}b we show the $log~L_{\ion{Mg}{ii}}$ versus
$M_V$ diagram for RS CVn stars and, for comparison, the regression
line applicable to normal stars (full line;
Eq. \ref{eq:lum-mv-normal}).  


To find an analytical representation to the data of RS CVn stars in
Fig. \ref{fig:magflunor}b, one is faced with the difficulty posed by
the very sparse distribution of the observations (see Table
\ref{tab:data2}), which would in fact lead to overweighting the few
most frequently observed stars.   For this reason we have
considered appropriate to perform a fit using the mean values for the
individual stars, weighted according to the observational errors
(i.e. the fixed 15\% percent error on the \ion{Mg}{ii} k fluxes, and
the parallax error; see Sect. \ref{sec:mg2fluxes}).


A linear fit to the data provides:

\begin{equation}
\it {log~L_{\ion{Mg}{ii}} = (31.35 {\pm} 0.03)- (0.32 {\pm} 0.01)~ M_V}
\label{eq:lum-mv-rs}
\end{equation}
\noindent  The fit is indicated as a dashed line in the above
figure. The corresponding correlation coefficient is r=0.90, to be
compared with r=0.82 if the individual measurements are taken into
account. The \ion{Mg}{ii} k line luminosity appears then to be
strongly correlated with $M_V$ (Eq. \ref{eq:lum-mv-rs}), contrary to
the \ion{Mg}{ii} line width (Sect. \ref{sec:wbrs}).

When comparing the linear fits of RS CVn stars with that of 
normal stars we notice the following important properties:\\

\noindent a) RS CVn stars are systematically brighter by ${\approx}$ 1
dex in the \ion{Mg}{ii} k line than normal stars. This  produces a
sort of luminosity gap (at a given $M_V$) between normal and RS CVn
stars.  \\

\noindent b) the slope of the two straight lines is the same within
the errors.\\


There are five objects that appear to be somewhat less luminous than
expected in the \ion{Mg}{ii} k line: RS CVn, Z Her, WW Dra, V815 Her, and
${\alpha}$~Aur.  Their positions are indicated with arrows in the
($L_{\ion{Mg}{ii}}$, $M_V$) plane of Fig. \ref{fig:magflunor}.  The
former three have a common peculiarity: the star that mainly
contributes to the \ion{Mg}{ii} k emission flux is the cool component,
while the residual flux at the bottom of the broad \ion{Mg}{ii}
photospheric absorption is mainly due to the hotter component.  Stars
with these characteristics were not included in the sample of RS CVn
stars by Montes et al. (1994).

As for ${\alpha}$~Aur, it cannot be considered as a typical RS CVn
star because, contrary to what generally observed in this class of
stars, the hotter component is more active than the cool one because
it is a faster rotator.  In the case of of the other deviating
star, V815 Her, its cool component is probably an M1--2 V star
according to CABS.  If this is correct, the object would better
qualify as a BY Dra star.

The diagrams in Fig. \ref{fig:magflunor} have a practical diagnostics
application: if the distance to a star is known (in addition to its
visual magnitude and \ion{Mg}{ii} k flux) one can easily distinguish a
quiet star from an active star such as RS CVn's.

\section{The \ion{Mg}{ii} k absolute luminosity vs. width relationship}
\label{sec:lummg2}
\begin{figure}
\includegraphics[width=8.8cm]{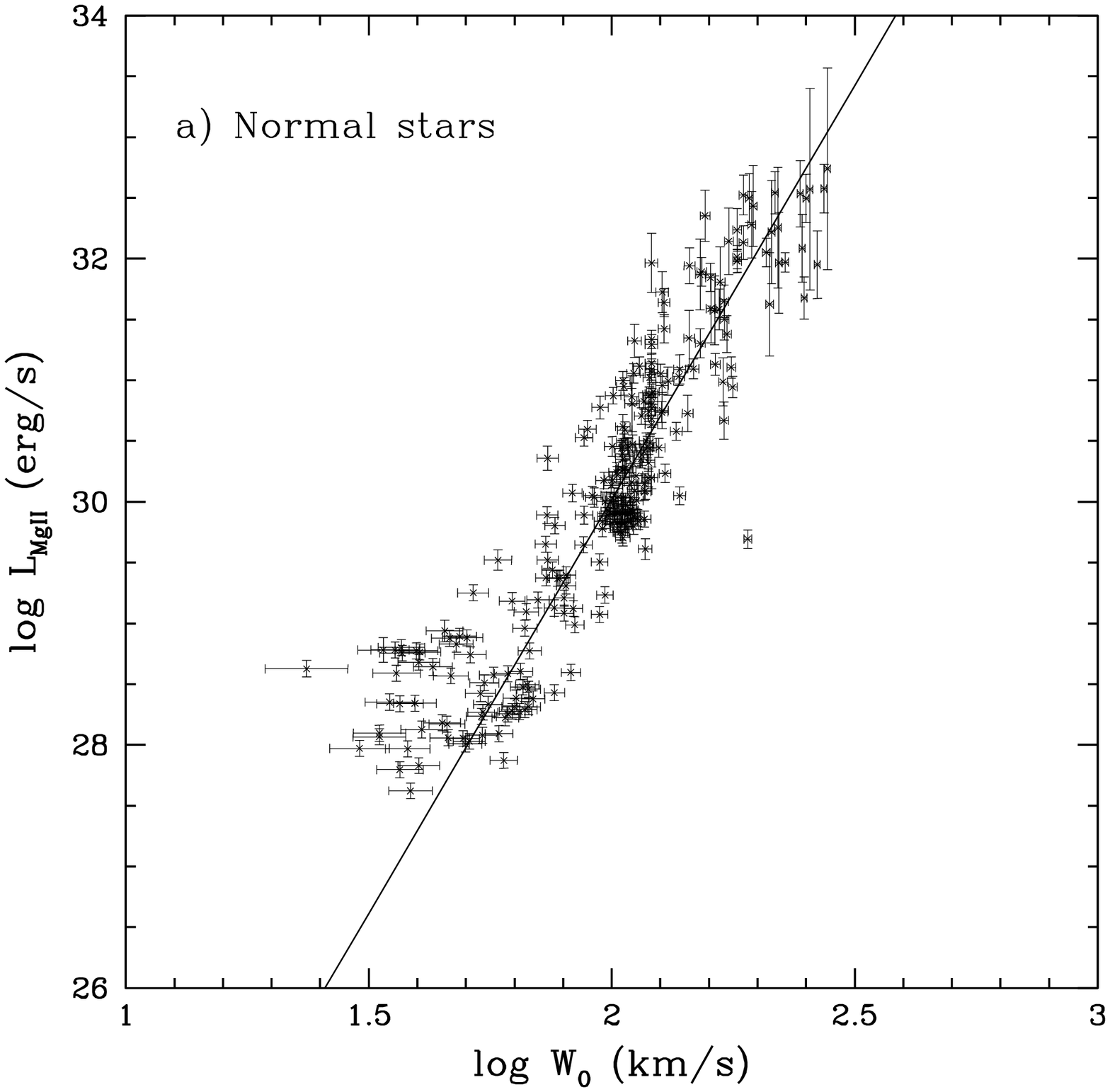}
\includegraphics[width=8.8cm]{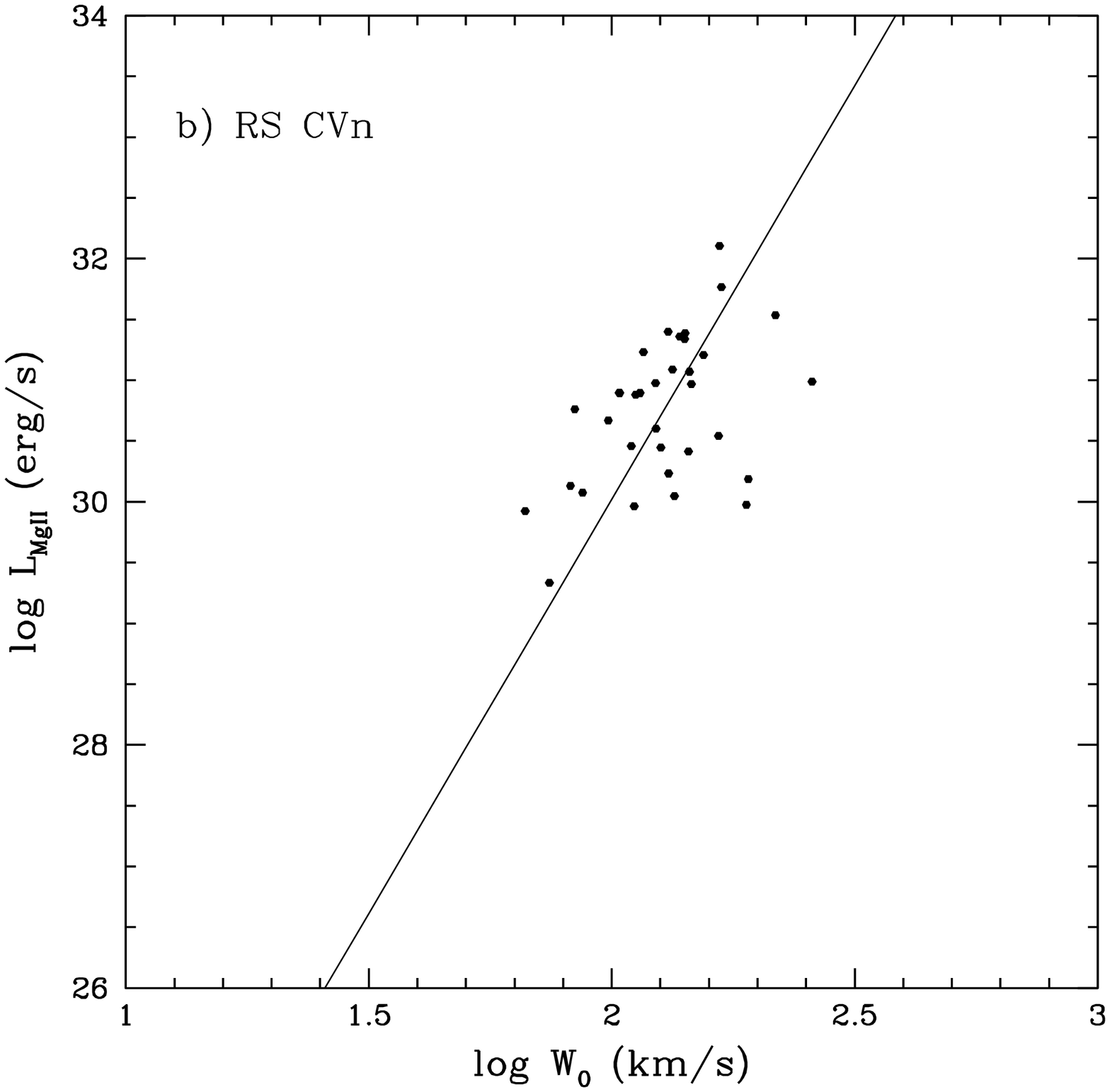}

\caption{The absolute luminosity in the \ion{Mg}{ii} k line as a
function of line width. Panel (a) gives the results for the sample of
normal stars in Paper I, together with the linear fit in
Eq. \ref{eq:widflueq}). Panel (b) gives the results for the RS CVn
stars in Table \ref{tab:data1} and shows, for comparison, the above
linear fit for normal stars}
\label{fig:lumwidth}
\end{figure}

In this Section we look for a correlation between the absolute
luminosity in the \ion{Mg}{ii} line and the corresponding line width
for the sample of normal stars in Paper I and, separately, for the
present sample of RS CVn stars.

\subsection{Normal Stars}
\label{sec:normal}

We have performed a linear fit to the $log~L_{\ion{Mg}{ii}}$ and
$log~W_{\circ}$ data for normal stars.  Taking the observational
errors into account we find:

\begin{equation}
\it {log~L_{\ion{Mg}{ii}} = (16.40 {\pm} 0.15)+ (6.81 {\pm} 0.07)~ log~W_{\circ}}
\label{eq:widflueq}
\end{equation}

\noindent where the errors on the coefficients represent the
confidence level of the regression line. The correlation coefficient
is r=0.9. We note that the above coefficients are close, but not
identical, to those obtained by combining Eqs. \ref{eq:bappueq} and
\ref{eq:lum-mv-normal}. This is an expected consequence of the
correlation coefficient not being unity.

In Fig. \ref{fig:lumwidth}a we plot the $log~L_{\ion{Mg}{ii}}$ and
$log W_{\circ}$ data together with the above linear relationship.  We
observe that the corresponding correlation coefficient is fairly high
in spite of the ${\approx}$ 6 orders of magnitude span in line
luminosity. This is a remarkable result in itself, which should
deserve proper study for its implications in the understanding of line
broadening mechanisms in the chromospheres of stars.

Another interesting aspect of Eq. \ref{eq:widflueq} is that it allows
a quite reliable distance determination to be made using
spectrophotometric data only.   The distance to the star is
readily obtained by comparing the observed \ion{Mg}{ii} k flux with
the absolute \ion{Mg}{ii} k luminosity corresponding, via
Eq. \ref{eq:widflueq}, to the observed value of $log W_{\circ}$.  If
proper account is taken for the spread of the data points about the
regression line, the standard error on the distance
determinations is about 32\%, to be compared with a value of about
25\% if the distances are obtained via the WB relationship in
Eq. \ref{eq:bappueq}.

This small difference in accuracy is probably due to the higher
accuracy of optical photometry compared to {\it IUE} photometry. We
conclude that Eq. \ref{eq:widflueq} represents a interesting
alternative to the classical Wilson--Bappu effect.
\subsection{RS CVn stars}
\label{sec:rrss}

Based on the data reported in Table \ref{tab:data2}, we plot in
Fig. \ref{fig:lumwidth}b, on a logarithmic scale, the mean \ion{Mg}{ii}
k absolute luminosity as a function of the mean line width for the RS
CVn stars in our sample.  

It clearly appears from the figure that the range of coordinate values
covered by RS CVn stars is quite narrow.  More importantly, the
correlation index corresponding to these data is extremely low: r=0.05
for individual measurements, and r= 0.47 for the mean values. This
prevents one to attempting any analytical representation to the data.

We then conclude that, at the present stage, the \ion{Mg}{ii} k line
cannot be used to determine the distance of RS CVn stars.

\section{Conclusions and discussion}
\label{sec:conclusions}
This paper is based on the analysis of the \ion{Mg}{ii} k emission
line width and flux in 34 RS CVn stars and 230 normal stars observed
by {\it IUE} at high resolution, for which {\it HIPPARCOS} parallax
determinations were available. The number of spectra used is very
large: 889 and 303, for the two classes, respectively.

\noindent Our results can be summarized as follows:
\begin{itemize}
\item The distribution of RS CVn stars in the ($log~{W_{\circ}}$,
$M_V$) plane does not show any regular pattern, as reflected by the
very low correlation index between these variables. Therefore, no
Wilson-Bappu relationship could be found for such stars, contrary to
the case of normal stars studied in Paper I. The absence of such a
correlation is due to their intrinsic variability and to the wide
spread of orbital and structural parameters of these binary systems
(see Sect. \ref{sec:wbrs}).
\item On the contrary, RS CVn stars are closely correlated in the
($M_V$, $log~L_{\ion{Mg}{ii}}$) plane, where they are well represented
by a straight line (Eq. \ref{eq:lum-mv-rs}). Such a strong correlation
is also found for normal stars, which lie along a nearly parallel line
(Eq. \ref{eq:lum-mv-normal}), but are systematically fainter in the
\ion{Mg}{ii} k line by a factor of ${\approx}$ 10 compared to RS CVn
stars (Figs. \ref{fig:magflunor}a,b).  These properties should
provide interesting constraints on the modeling of RS CVn
chromospheres and represent a reliable tool to discriminate between
normal and RS CVn stars, at least in the range of $M_V$ covered
(${\approx}$ 5 to -2).

\item The data for normal stars are strongly correlated in the
($log~W_{\circ}$, $log~L_{\ion{Mg}{ii}}$) plane (see Fig.
\ref{fig:lumwidth}a), allowing the linear relationship in
Eq. \ref{eq:widflueq} to be defined.  This equation represents an
interesting alternative to the standard Wilson--Bappu method for
distance determinations, it offers nearly the same accuracy, is based
only on spectrophotometric data of the \ion{Mg}{ii} line, and is
applicable to stars covering $\approx$ 6 orders of magnitude in line
luminosity.\\ 
On the contrary, no similar correlation has been found
for RS CVn stars (Fig. \ref{fig:lumwidth}b).

\end{itemize}

\begin{acknowledgements}
The authors are grateful to Prof. J. Linsky  for stimulating
suggestions and to Dr. D.  Marinucci for useful comments.

\end{acknowledgements}

\end{document}